\documentclass[fleqn,usenatbib]{mnras}

\usepackage{newtxtext,newtxmath}

\usepackage[T1]{fontenc}

\DeclareRobustCommand{\VAN}[3]{#2}
\let\VANthebibliography\thebibliography
\def\thebibliography{\DeclareRobustCommand{\VAN}[3]{##3}\VANthebibliography}

\usepackage{graphicx}	
\usepackage{amsmath}	

\usepackage{hyperref}
\hypersetup{
    colorlinks=true,
    linkcolor=cyan,
    citecolor=cyan,
    urlcolor=cyan,
    }

\usepackage{orcidlink}

\title[Blending-induced beating and emission in the symbiotic star Terz V 2513]{Blending-induced beating and emission in the symbiotic star Terz V 2513
}

\author[J. Merc et al.]{
J. Merc,$^{1,2}$\thanks{E-mail: jaroslav.merc@mff.cuni.cz}\orcidlink{0000-0001-6355-2468}
J.~Miko\l{}ajewska,$^{3}$\orcidlink{0000-0003-3457-0020}
C.~Ga\l{}an,$^{3}$\orcidlink{0000-0002-7868-5017}
K.~I\l{}kiewicz,$^{3}$\orcidlink{0000-0002-4005-5095}
P.~G.~Beck,$^{2,4}$\orcidlink{0000-0003-4745-2242}
B.~Monard,$^{5}$\orcidlink{0009-0004-6582-0787}
M.~Gromadzki$^{6}$\orcidlink{0000-0002-1650-1518}
\\
$^{1}$Astronomical Institute, Faculty of Mathematics and Physics, Charles University, V Hole\v{s}ovi\v{c}k{\'a}ch 2, 180 00 Prague, Czech Republic\\
$^{2}$Instituto de Astrof\'isica de Canarias, Calle Vía Láctea, s/n, E-38205 La Laguna, Tenerife, Spain\\
$^{3}$Nicolaus Copernicus Astronomical Center, Polish Academy of Sciences, Bartycka 18, 00–716 Warsaw, Poland\\
$^{4}$Departamento de Astrof\'{\i}sica, Universidad de La Laguna, E-38206 La Laguna, Tenerife, Spain\\
$^{5}$Kleinkaroo Observatory, Sint Helena 1B, PO Box 281, Calitzdorp 6660, South Africa\\
$^{6}$Astronomical Observatory, University of Warsaw, Al. Ujazdowskie 4, 00-478 Warszawa, Poland
}

\date{Accepted 2025 November 20. Received 2025 November 14; in original form 2025 August 27}

\pubyear{\the\year{}}

\begin{document}
\label{firstpage}
\pagerange{\pageref{firstpage}--\pageref{lastpage}}
\maketitle

\begin{abstract}
We present a detailed analysis of Terz V 2513 (=2MASS J17334728-2719266), a poorly studied symbiotic star. Our motivation was a peculiar beating pattern in its light curves from all-sky surveys and our own observations. Using \textit{Gaia} DR3 and OGLE-IV photometry, we show that this variability arises from blending with a nearby, unrelated Mira variable (\textit{Gaia}
DR3 406134544052580377 = OGLE-BLG-LPV-241930). Analysis of VPHAS+ and Pan-STARRS imaging, combined with optical and infrared spectroscopy from the Southern African Large Telescope and ESO New Technology Telescope, further reveals that the symbiotic star has been misidentified in the literature. We identify the correct counterpart as \textit{Gaia} DR3 4061345440488592896 (=OGLE-BLG-LPV-241932), a Mira with a 161-day period. Its infrared spectrum displays prominent emission lines and is remarkably similar to those of other symbiotic Miras. Based on our data and previous studies, Terz V 2513 likely experienced a symbiotic nova outburst in the past. This study highlights the importance of careful analysis of survey light curves in crowded fields and demonstrates how combining multi-wavelength photometry, spectroscopy, and high-precision \textit{Gaia} data can disentangle blended sources and accurately determine their nature.
\end{abstract}




\section{Introduction}

Symbiotic stars, binaries composed of a cool red giant donor and a compact accretor (typically a white dwarf or neutron star), exhibit a wide range of photometric and spectroscopic variability on timescales ranging from minutes to decades \citep[see, e.g., ][]{2012BaltA..21....5M,2019arXiv190901389M,2025Galax..13...49M}. This variability arises from a combination of processes, including mass transfer between the components, intrinsic variability of the red giant, and orbital motion. For instance, the pulsation periods of the giants typically span tens to hundreds of days, while the orbital periods are also on the order of hundreds of days, often overlapping \citep[e.g.,][]{2009AcA....59..169G,2013AcA....63..405G,2019RNAAS...3...28M,2019AN....340..598M}. Proper analysis of such variability requires long-term light curves spanning several years. Today, these are frequently obtained from all-sky surveys, which provide readily accessible and extensive datasets.

While such surveys have significantly advanced the study of symbiotic variability, they also present challenges. The absence of color information can hinder efforts to distinguish between different physical origins of the observed variability. Additionally, the relatively large pixel scales of many surveys can lead to blending and contamination, especially in crowded fields, further complicating the interpretation of light curves.

In this work, we investigate one such case, identified during our analysis of photometric variability of a sample of symbiotic stars discovered with the Southern African Large Telescope (SALT), as presented by \citet{2014MNRAS.440.1410M} and \citet{Merc+SALT_PaperII}. The object in question is Terz V 2513 (=2MASS J17334728-2719266), which was classified as an S-type symbiotic star with an M2 giant donor. Its optical spectrum shows emission lines of \ion{H}{i}, \ion{He}{i}, \ion{He}{ii}, and [\ion{O}{iii}]. \citet{2014MNRAS.440.1410M} also suggested that the system may have undergone an outburst in the past. Aside from its initial classification and brief discussion, no further detailed investigation of the star has been published.

We analyze available information in this work. Section~\ref{sec:observations} describes the photometric and spectroscopic datasets employed. In Section~\ref{sec:results}, we analyze the field of Terz V 2513, examine the spectral energy distributions of two stars in the vicinity of the infrared source, and present both the optical SALT spectrum and the infrared spectra obtained with the ESO New Technology Telescope (NTT), alongside an investigation of the photometric variability. Section~\ref{sec:discussion} discusses the symbiotic classification and nature of the system, and Section~\ref{sec:conclusions} summarizes our findings.

\section{Observational data}\label{sec:observations}

\subsection{Photometry}
Our time series were obtained in the $V$ and $I$ filters using the 35 cm Schmidt–Cassegrain and Ritchey-Chretien telescopes equipped with SBIG CCD cameras at the Kleinkaroo Observatory in South Africa. The observations cover the interval starting from June 2015.

Additional data are available in the $o$- and $c$-band from the Asteroid Terrestrial-impact Last Alert System (ATLAS) project \citep{2018PASP..130f4505T,2020PASP..132h5002S}, obtained via the ATLAS Forced Photometry server \citep{2021TNSAN...7....1S}, the Optical Gravitational Lensing Experiment
(OGLE) IV survey \citep{2015AcA....65....1U} in $I$-band, in $g$ and $r$ from the Zwicky Transient Facility survey \citep[ZTF;][]{2019PASP..131a8003M}, and in the $V$ and $g$ bands from the All-Sky Automated Survey for Supernovae \citep[ASAS-SN;][]{2014ApJ...788...48S,2017PASP..129j4502K}. Part of the light curve is also covered by \textit{Gaia} DR3 epoch photometry \citep[$BP, G, RP$ filters;][]{2023A&A...674A...1G}.

To search for periodicities in the light curves, we performed period analysis using the Lomb-Scargle method \citep{1976Ap&SS..39..447L,1982ApJ...263..835S}  implemented in the {\tt astropy} Python package \citep{2013A&A...558A..33A,2018AJ....156..123A,2022ApJ...935..167A}. 

We complemented our analysis of field of the target star and spectral energy distribution with imaging from several large-scale surveys, namely the Panoramic Survey Telescope and Rapid Response System \citep[Pan-STARRS;][]{2012ApJ...750...99T,2016arXiv161205560C,2020ApJS..251....7F}, the VST Photometric H$\alpha$ Survey of the Southern Galactic Plane and Bulge \citep[VPHAS+;][]{2014MNRAS.440.2036D}, the VISTA Variables in the Vía Láctea \citep[VVV;][]{2012A&A...537A.107S}, and the DECam Plane Survey 2 \citep[DECaPS2;][]{2023ApJS..264...28S}.

\subsection{Spectroscopy}
The optical spectrum of the target was obtained on September 10, 2013, using the Robert Stobie Spectrograph \citep[RSS;][]{2003SPIE.4841.1463B,2003SPIE.4841.1634K} on the SALT telescope \citep{2006SPIE.6267E..0ZB,2006MNRAS.372..151O} under program 2013-1-RSA\_POL-001 (PI: Miszalski) and was already presented in \citet{2014MNRAS.440.1410M}. Data were obtained in single RSS configuration with the PG900 grating and a 1.5\arcsec{} slit at a positional angle of 0\textdegree{}, providing a wavelength coverage of $\sim$4\,340--7\,420 \AA{} and a resolution of 6.2 \AA{} (see more details in the aforementioned work).

The near-infrared spectra used in this work were obtained under programs 097.D-0338 and 099.D-0647 (PI: Ga\l{}an) on June 16, 2016 (Terz V 2513 field), June 19, 2016 (JaSt 79), and July 10, 2017 (V5590 Sgr) with SofI spectrograph \citep{1998Msngr..91....9M} on the ESO NTT (3.58m) in low-resolution mode (R$\sim$600–950) using slit 0.6", and Blue (0.95–1.64 $\mu$m) and Red (1.53–2.52 $\mu$m) grism. 

The spectra were processed using standard methods with IRAF\footnote{IRAF is distributed by the National Optical Astronomy Observatories, which are operated by the Association of Universities for Research in Astronomy, Inc., under cooperative agreement with the National Science Foundation.} packages. To make it possible to properly correct data on the effects of crosstalk, biases, dark current, and background strong in infrared, for each spectral range, four frames were taken in the ABBA sequence. The data processing included additional correction on the bad pixels, flat-field effect, and wavelength calibration (using the spectra of lines from a xenon lamp). The spectra were corrected for telluric lines by the reference to hot B-type standard stars \citep[][]{2003AJ....125.2645C} using the synthetic models BT-NextGen, and by the use of synthetic spectra of the atmosphere in La Silla, generated using the TAPAS service\footnote{Transmissions Atmosphériques Personnalisées Pour l’Astronomie; \url{http://ether.ipsl.jussieu.fr/tapas/}} \citep{2014A&A...564A..46B}, and convolved with a rotational and instrumental profile.  The final relative flux calibration was performed using the ESO library of stellar spectra for spectrophotometric calibration\footnote{\url{https://www.eso.org/sci/observing/tools/standards/IR_spectral_library.html}} \citep{1998PASP..110..863P}.
\begin{figure}
\includegraphics[width=0.97\columnwidth]{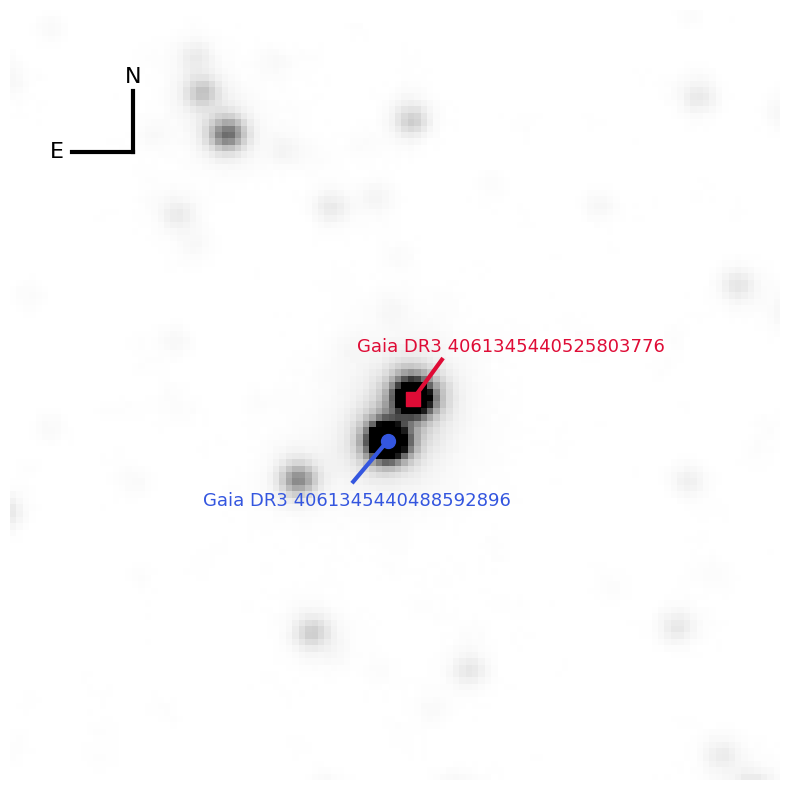}
    \caption{Pan-STARRS $r$-band image of the field around Terz V 2513. The displayed field of view is 30\arcsec$\times$30\arcsec. The positions of the two stars discussed in the text are indicated.}
    \label{fig:panstarrs}
\end{figure}

\begin{figure}
\includegraphics[width=0.97\columnwidth]{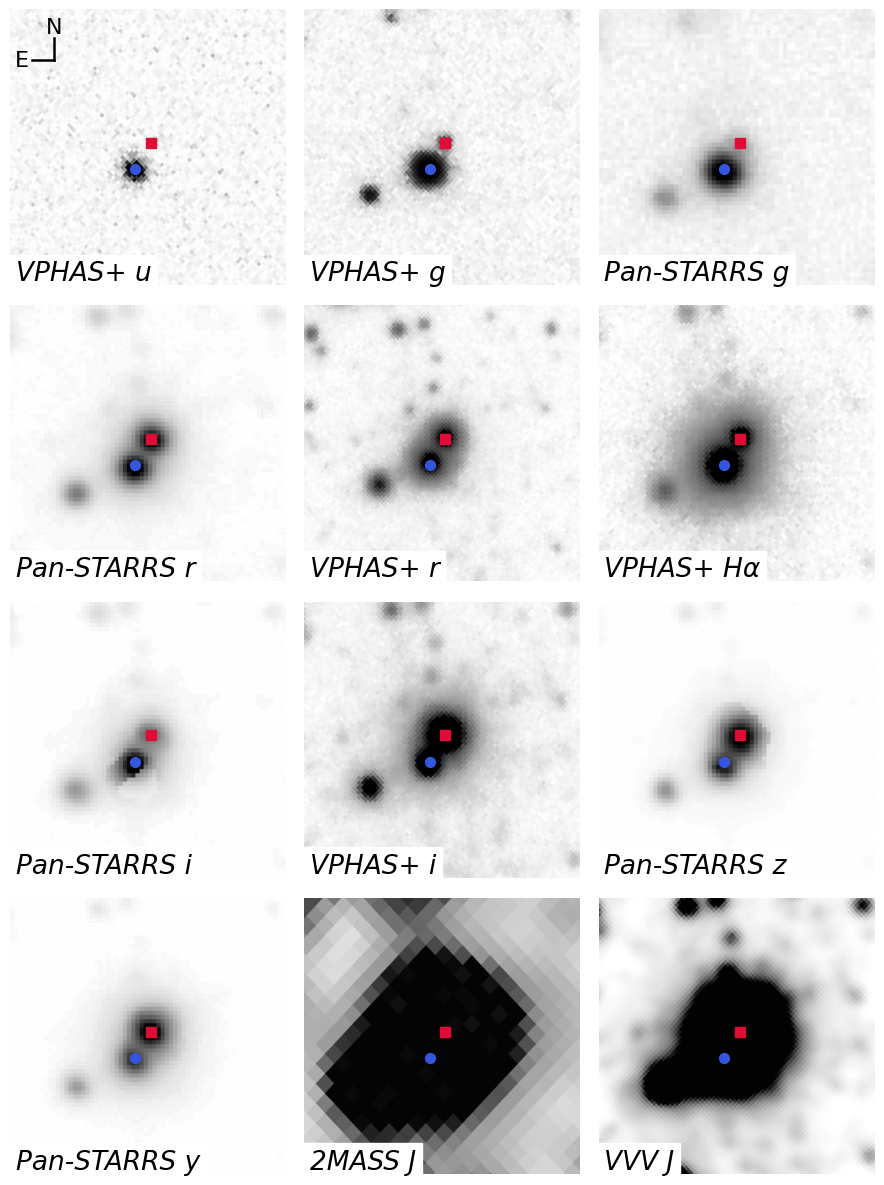}
    \caption{Multi-band images of the field around Terz V 2513. Images from VPHAS+ (SDSS $u$, SDSS $g$, SDSS $r$, H$\alpha$, SDSS $i$ bands), Pan-STARRS ($g$, $r$, $i$, $z$, $y$ bands), 2MASS $J$, and VVV $J$-band are shown. The displayed field of view is $\sim$16.5\arcsec$\times$16.5\arcsec. The positions of the two stars are indicated in the same way as in Fig. \ref{fig:panstarrs}. Images are arranged in order of increasing filter effective wavelength.}
    \label{fig:vphas}
\end{figure}

\begin{figure*}
\includegraphics[width=\textwidth]{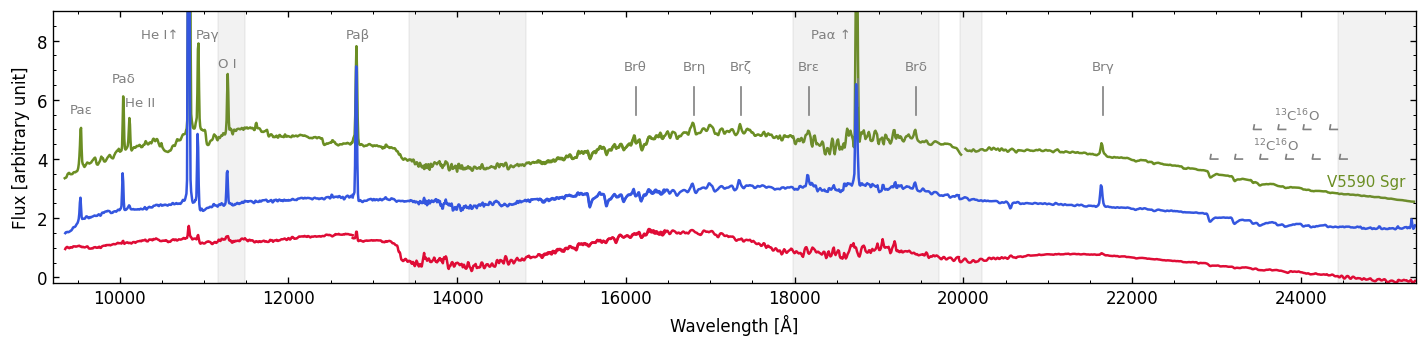}
    \caption{NTT/SofI spectra of both sources. Line colors correspond to those of the respective stars in Fig. \ref{fig:panstarrs}. For comparison, the spectrum of the symbiotic nova with a Mira-type donor V5590 Sgr (Nova Sgr 2012b) is shown in green. Major emission and absorption features are labeled in gray. Shaded regions indicate wavelength intervals most affected by telluric absorption.}
    \label{fig:sofi_spec}
\end{figure*}

\begin{figure}
\includegraphics[width=\columnwidth]{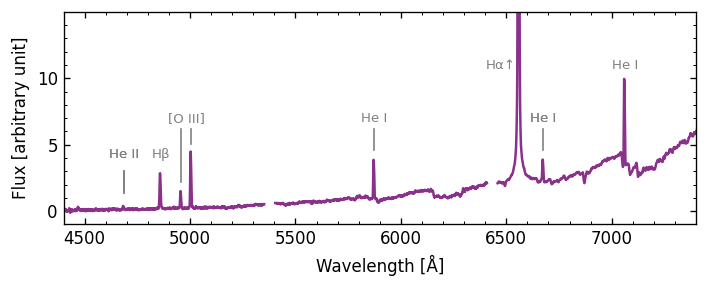}
    \caption{SALT/RSS spectrum of Terz V 2513. Light from both sources visible in Fig. \ref{fig:panstarrs} is blended in this spectrum, and is represented here by the purple color. The spectrum is not dereddened. Due to limitations in absolute flux calibration arising from the moving pupil design of SALT, the flux is normalized to the median value. The identification of emission lines is shown in gray.}
    \label{fig:salt_spec}
\end{figure}

\begin{figure}
\includegraphics[width=\columnwidth]{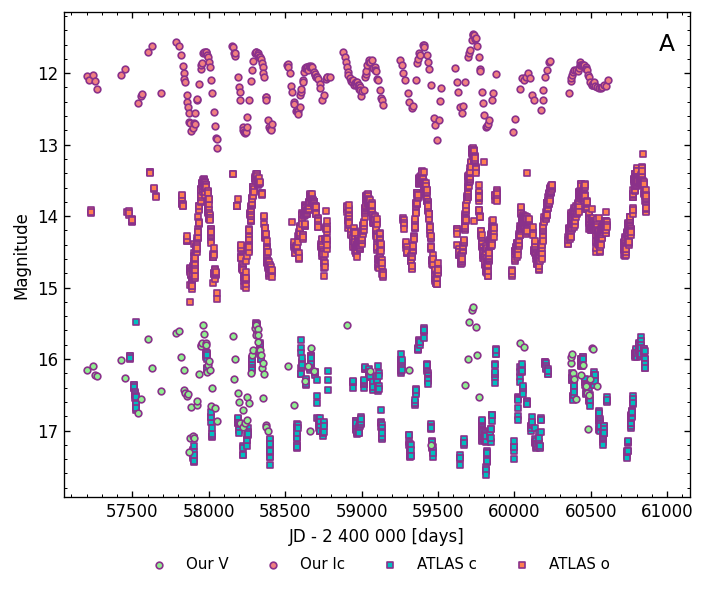}
\includegraphics[width=\columnwidth]{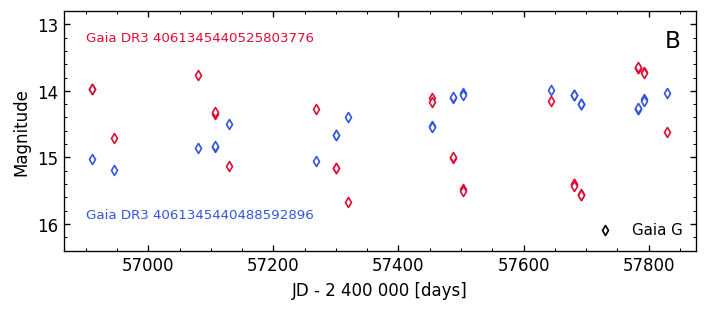}
\includegraphics[width=\columnwidth]{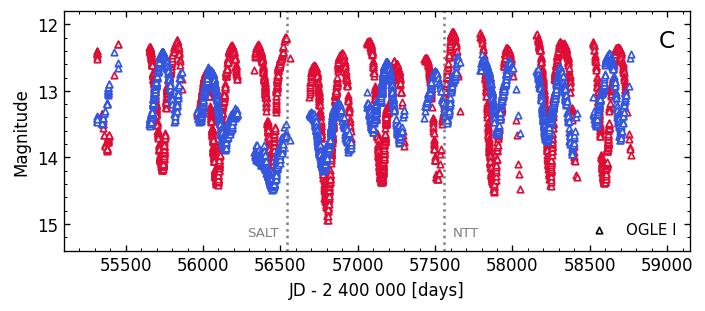}
    \caption{Photometry of the studied sources. \textbf{A:} Light curves in the $V$ and $I_c$ bands from our observations, along with ATLAS $o$ and $c$ bands. Light from both stars is blended in these data (indicated by the purple symbol borders). \textbf{B:} \textit{Gaia} DR3 $G$-band light curves of the two sources. Point colors correspond to the star colors in Fig. \ref{fig:panstarrs}. \textbf{C:} OGLE-IV $I$-band light curves of both sources. The epochs of the SALT and NTT spectra are marked by vertical lines. Note the different JD ranges shown on the x-axes of each panel.}
    \label{fig:photometry}
\end{figure}

\section{Results}\label{sec:results}

\subsection{Field of Terz V 2513}
Terz V 2513 was first selected as a symbiotic candidate by \citet{2014MNRAS.440.1410M} based on its appearance in the AAO/UKST SuperCOSMOS H$\alpha$ Survey \citep[SHS;][]{2005MNRAS.362..689P} and the Two Micron All-Sky Survey \citep[2MASS;][]{2006AJ....131.1163S}. In both surveys, only a single source was resolved within a few arcseconds, with the SHS H$\alpha$ emitter matched to 2MASS~J17334728-2719266.

Our closer inspection, however, reveals that near the 2MASS position there are two similarly bright stars, separated by 2.01 arcsec according to \textit{Gaia} DR3. Both are resolved in Pan-STARRS imaging (at least in most epochs), see Fig. \ref{fig:panstarrs}, where their \textit{Gaia} identifications are indicated. The northern star (marked in red; Gaia DR3 4061345440525803776) lies 0.52 arcsec from the 2MASS coordinates, while the southern star (marked in blue; Gaia DR3 4061345440488592896) is offset by 1.50 arcsec.

The current symbiotic star catalogs \citep[e.g., the \textit{New Online Database of Symbiotic Variables};][]{2019RNAAS...3...28M,2019AN....340..598M,Merc+NODSV2025} associate the object with the \textit{Gaia} ID of the northern source, consistent with its proximity to the 2MASS position. The same association appears in the \textit{Gaia} DR3 variability tables (specifically \textit{vari\_classifier\_result}), where both stars are flagged as variable: the northern star is classified as a symbiotic system, while the southern is classified as a long-period variable \citep[see][]{2023A&A...674A..14R,2023A&A...674A..13E}. The coordinates used for this cross-match originate from the catalog of \citet{2019ApJS..240...21A}.

\subsection{Multi-color photometry and spectra}
Figure~\ref{fig:vphas} shows multi-band images of the field of TerzV2513 from Pan-STARRS (\textit{g}, \textit{r}, \textit{i}, \textit{z}, \textit{y} filters), VPHAS+ (\textit{u}, \textit{g}, \textit{r}, H$\alpha$, \textit{i} filters), as well as 2MASS and VVV \textit{J}-band images. The imaging clearly shows no blue excess associated with the northern star, whereas the southern star exhibits a strong blue excess, evident in the VPHAS+ \textit{u} and \textit{g} images and in the Pan-STARRS \textit{g} band.

In \textit{Gaia} DR3, both stars have comparable \textit{G}-band magnitudes (14.85 and 14.42 mag for the northern and southern sources, respectively), but the southern star is noticeably bluer (\textit{BP$-$RP} = 2.99 mag) than the northern one (\textit{BP$-$RP} = 4.08 mag). The same trend is seen in DECaPS~DR2 photometry, with the southern star being significantly bluer.

The southern source is also conspicuously brighter in the VPHAS+ H$\alpha$ filter. This is consistent with the \textit{Gaia} detection of H$\alpha$ emission from this star and with the pseudo-equivalent width measured from its BP/RP low-resolution spectrum, as published in \textit{Gaia} DR3.

Our NTT/SofI spectra (Fig. \ref{fig:sofi_spec}) clearly resolve the two components. They confirm that the southern star is exhibiting the strong emission-line spectrum, displaying prominent \ion{H}{i} Paschen and Brackett lines, along with \ion{O}{i}, \ion{He}{i}, and faint \ion{He}{ii} emission. The continuum also reveals strong molecular absorption features, most notably from CO bands, characteristic of a cool giant donor. The spectrum of northern star shows only very faint emission.

The optical spectrum obtained with SALT by \citet{2014MNRAS.440.1410M}, which led to the symbiotic classification, shows an M-type continuum with emission lines of \ion{H}{i}, \ion{He}{i}, [\ion{O}{iii}], and \ion{He}{ii} (Fig. \ref{fig:salt_spec}). However, those observations did not resolve the two stars, making it impossible to assign the individual optical absorption bands to either component. The spectrum was obtained at a time when the southern star was fainter than the northern one (see Fig.~\ref{fig:photometry}), but this does not affect any of the conclusions of this work. In the following, data for the individual stars are shown in red (northern) and blue (southern), while blended measurements are shown in purple.

\begin{figure}
\includegraphics[width=\columnwidth]{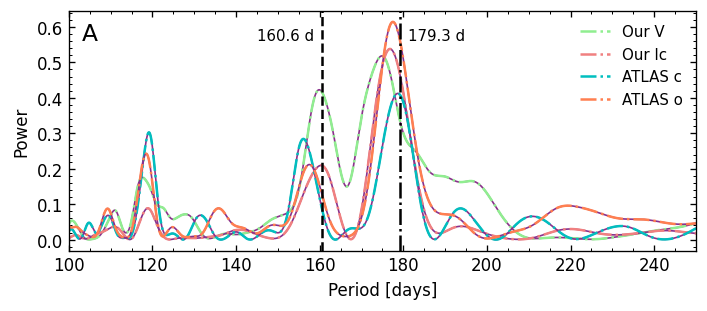}
\includegraphics[width=\columnwidth]{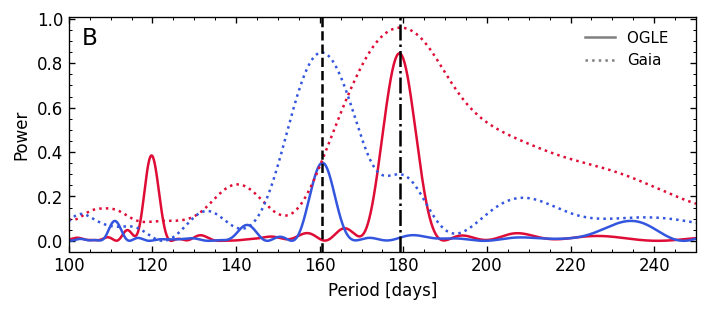}
    \caption{Lomb-Scargle periodograms of the photometric data. \textbf{A:} Periodograms based on our observations and ATLAS data, in which the two sources are blended. Colors match those in Fig. \ref{fig:photometry}. Vertical lines indicate prominent periods identified from OGLE data of the individual stars. Additional peaks near 111 and 120 days are yearly aliases of the true periods. \textbf{B:}~Periodograms derived from \textit{Gaia} (dotted lines) and OGLE (solid lines) data, where the sources are well separated. Colors correspond to those of the stars in Fig. \ref{fig:panstarrs}.}
    \label{fig:periodogram}
\end{figure}

\subsection{Variability}
We have been monitoring the system from South Africa since June 2015 in the $V$ and $I$ bands. Our light curves, together with ATLAS data, are shown in Fig. \ref{fig:photometry}A. At first glance, the variability amplitude changes over time, producing a characteristic beating pattern. Period analysis of these datasets (Fig. \ref{fig:periodogram}A) consistently reveals two prominent periods of about 160 and 180 days in all blended datasets. Since both sources are unresolved in these photometric observations, these data alone cannot determine whether both periodicities originate from the same star or each is associated with a different one.

This ambiguity is resolved thanks to higher-angular-resolution photometry from the \textit{Gaia} satellite ($G$-band light curves are shown in Fig. \ref{fig:photometry}B) and from the OGLE-IV survey in the $I$ band (Fig. \ref{fig:photometry}C). These data are available for each source separately and confirm that the shorter period ($P=160.6$~d) belongs to the southern star, while the longer period ($P=179.3$ d) is exhibited by the northern star (see Fig. \ref{fig:periodogram}B). The observed beating pattern in blended data is therefore entirely due to the combination of these two periodic, but unrelated signals. The southern star also shows hints of an additional long-term variability (see the drop in the average brightness of the star around JD 2\,456\,500, visible in the OLGE $I$ band; Fig. \ref{fig:photometry}C), but the current dataset is insufficient to constrain its period or nature.

For completeness, we also present light curves from ASAS-SN and ZTF in Fig.~\ref{fig:phot_app}A,B. The ASAS-SN data, due to the faintness of the system, are of lower quality but still show similar variability patterns. In the $V$ band, the periodogram shows a single peak between the two main periods (Fig. \ref{fig:phot_app}C), whereas the $g$~band data suggest two periods close to the OGLE values. In ZTF $g$~band, the shorter period dominates (unsurprising, as the southern star is brighter in blue than the northern one). The shorter period is also clearly detectable in $r$ band, where the longer period (around 190 days) is additionally seen.

\section{Discussion} \label{sec:discussion}
There are two similarly bright stars in the immediate vicinity of 2MASS J17334728-2719266, an object classified as a symbiotic variable by \citet{2014MNRAS.440.1410M}. There is no evidence in our data that the two sources are physically related. Both stars are variable, exhibiting Mira-like pulsations with close periods of 160.6 days (southern star) and 179.3 days (northern star). They are already listed in OGLE-IV as Mira variables \citep[][]{2022ApJS..260...46I}:
the southern star as OGLE-BLG-LPV-241932, $P=160$~d, $A=0.97$~mag, and the northern star as OGLE-BLG-LPV-241930, $P=179$~d, $A=1.89$~mag. These parameters are in excellent agreement with the results of our period analysis.

Both stars have negative parallaxes in \textit{Gaia}~DR3 ($-0.0058\pm0.0658$~mas for the southern star, and $-0.2866\pm0.1291$~mas for the northern), which prevents a direct distance determination or a reliable placement on the \textit{Gaia} color–magnitude diagram \citep[like the one of][]{2025A&A...696A.243G}. Nevertheless, their Mira pulsations indicate that both are evolved cool giants. 

We estimated their absolute $K$-band magnitudes using the period–luminosity relation for oxygen-rich Miras. The calibration of \citet{2008MNRAS.386..313W} yields $M_K=-6.54$ and $-6.71$~mag for the northern and southern star, respectively. Using the more recent calibration of \citet{2023MNRAS.523.2369S} gives similar results: $M_K=-6.45$ and $-6.65$~mag.

\subsection{Symbiotic identification}
There is compelling evidence that the true symbiotic source is the southern star, contrary to the current association of the symbiotic classification with the northern source. As shown above, the blue excess expected from an accreting hot companion is clearly associated with the southern star.

The VPHAS+ images are available from only one epoch, but fortunately this corresponds to a time when both stars were varying almost synchronously, with overlapping minima and maxima (around JD 2\,456\,500). This means that the difference in short-wavelength brightness between the two stars cannot be attributed to different pulsation phases. Stacks from multiple Pan-STARRS epochs show the same result. The only case where the northern star appears brighter than the southern in the $g$ band occurs around JD 2\,455\,822, when the northern star was near its brightness maximum and the southern star near its minimum. In all other Pan-STARRS epochs, the southern star is brighter in $g$. This trend is confirmed by median \textit{Gaia} DR3 $BP-RP$ colors, as well as by DECaPS DR2 data taken at epoch with minimal phase offset between the pulsations of the two stars.

Further evidence comes from the detection of H$\alpha$ emission: \textit{Gaia}~DR3 reports it only for the southern star, in agreement with VPHAS+ H$\alpha$ imaging. Our NTT/SofI spectra also confirm that the southern star is the emission-line object. The very faint emission seen in the spectrum of the northern star is likely an artifact of data processing, although we cannot exclude the possibility of faint extended nebulosity from the southern star contaminating its spectrum.

The fact that the 2MASS position lies closer to the northern star may partly be due to variability, as 2MASS observations are from a single epoch. Moreover, if both stars have comparable $G$-band magnitudes and the flux of the southern star comes from both a cool Mira and a nebula and/or hot companion, while the northern star is presumably single, then the contribution of the red Mira in the northern star relative to Mira in the southern must be stronger. For similar absolute magnitudes, this would imply the northern star is closer, and thus also brighter in the infrared, shifting the center of blended light toward itself.

\subsection{Outburst activity of Terz V 2513} 
Based on the large difference in brightness between recent observations of Terz V 2513 and those reported by \citet{1991A&AS...90..451T}, \citet{2014MNRAS.440.1410M} proposed that the star had undergone either a Z And-type outburst or a symbiotic nova eruption. The photographic plates analyzed by \citet{1991A&AS...90..451T} show a maximum brightness of $R < 8$ mag on September 15, 1987, and a minimum brightness of $\sim$18 mag on July 29, 1968. Variability was also detected in the $B$ and $V$ bands, although no magnitudes were reported. The exposure times (20--30 minutes) used to reach a limiting magnitude of $R \sim 18$--19.5 mag imply that the reported maximum brightness may have been affected by saturation and could in fact have been even higher. Nevertheless, the data clearly indicate a very large photometric amplitude in the optical. Such behavior is inconsistent with the relatively modest brightenings seen in Z And-type outbursts and instead strongly points toward a symbiotic nova eruption. Unfortunately, the duration of this event cannot be fully constrained. \citet{1991A&AS...90..451T} observed the field several times prior to September 15, 1987, with the last pre-outburst plate obtained on August 18 of the same year. However, they reported only the epochs of maximum and minimum brightness, making it impossible to reconstruct the intermediate behavior. For instance, it remains unclear whether the star could not have been only slightly fainter on the plates collected earlier in 1987. On the other hand, given the presence of a nearby star close to Terz V 2513, any magnitudes reported in other catalogs, such as the $R$ magnitude from USNO-A2.0 \citep[][]{1998usno.book.....M}, are likely significantly contaminated by this neighbor outside the outburst maximum.

\begin{figure}
\includegraphics[width=\columnwidth]{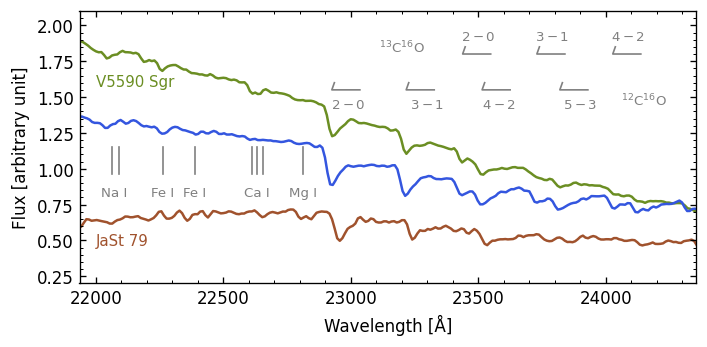}
    \caption{Part of the NTT/SofI spectrum of the symbiotic source (blue), compared with the spectra of the symbiotic nova V5590~Sgr (Nova Sgr 2012b; green), which hosts a Mira-type donor, and the symbiotic Mira JaSt 79 (brown). The most prominent absorption features are indicated in grey.}
    \label{fig:sofi_spec_zoom}
\end{figure}

Our photometric data further confirm that Terz V 2513 hosts a Mira-type pulsating donor, reinforcing the nova interpretation, since Z And-type outbursts are rarely associated with symbiotic Miras. Only a few such cases are known (e.g., RX Pup; V407 Cyg; \citealp{2002AIPC..637...42M,2007BaltA..16...23S,2013AN....334..810E}), and notably, in the catalogue of \citet{2000A&AS..146..407B}, which also lists outburst characteristics of symbiotic stars, no such eruption is associated with any symbiotic Mira. In Fig. \ref{fig:sofi_spec}, we compare the infrared spectrum of Terz V 2513 with that of the symbiotic nova V5590 Sgr (Nova Sgr 2012b), which also hosts a Mira-type donor \citep{2009A&A...496..813M,2014MNRAS.443..784M}. Both spectra, obtained within the same observational program, show remarkable similarities. A closer view of the $K$-band region (Fig.~\ref{fig:sofi_spec_zoom}) reveals a wealth of CO molecular bands. For reference, we also include the spectrum of another confirmed symbiotic Mira, JaSt 79 \citep{2013MNRAS.432.3186M}, likewise observed in the same program, which closely resembles that of Terz V 2513, and exhibits similar absorption features (e.g., CO bands) originating from the Mira photosphere.

The infrared spectroscopy therefore provides additional strong evidence for the symbiotic nature of the southern star, and taken together, the photometric and spectroscopic characteristics indicate that Terz V 2513 is likely a post-nova symbiotic system.

\section{Conclusions}\label{sec:conclusions}

We have analysed new and archival photometric observations, imaging, and optical/infrared spectroscopy of Terz V 2513, a poorly studied symbiotic star originally discovered by \citet{2014MNRAS.440.1410M}. Our main motivation was the peculiar beating pattern in our light curve of this system, also evident in ASAS-SN and ATLAS data.

Using \textit{Gaia} DR3 and OGLE-IV photometry, we identified the origin of this behavior: the light of Terz V 2513 is blended with that of a second, unrelated Mira variable. The two stars have similar pulsation periods (161 and 180 days), producing the observed beating.

Analysis of VPHAS+ and Pan-STARRS imaging, \textit{Gaia} DR3 data, and especially the infrared spectra of both Miras, reveals that the symbiotic star has been misidentified in the literature. We show that the correct counterpart is \textit{Gaia} DR3 4061345440488592896 (=OGLE-BLG-LPV-241932), which exhibits a pulsation period of 161 days. The infrared spectrum of this star shows strong emission lines and is strikingly similar to those of other symbiotic Miras, such as JaSt 79 and V5590 Sgr. The latter underwent a symbiotic nova outburst, and it is likely that Terz V~2513 experienced a similar event.

This case underscores the importance of careful analysis of survey light curves, particularly in crowded fields where blending can significantly affect the interpretation of variability. It also demonstrates how the combination of multi-wavelength photometry, spectroscopy, and high-precision data from \textit{Gaia} can be used to disentangle blended sources and securely identify the true nature of variable stars.

\section*{Acknowledgements}

We thank the anonymous referee for the careful review and helpful suggestions that improved the manuscript. The research of JaM was supported by the Czech Science Foundation (GACR) project no. 24-10608O. JMik was supported by the Polish National Science Centre (NCN) grant
2023/48/Q/ST9/00138. KI was supported by the Polish NCN grant 2024/55/D/ST9/01713. PGB acknowledges support by the Spanish Ministry of Science and Innovation with the \textit{Ram{\'o}n\,y\,Cajal} fellowship number RYC-2021-033137-I and the number MRR4032204. PGB and JaM acknowledge support from the Spanish Ministry of Science and Innovation with the 
proyecto plan nacional \textit{PLAtoSOnG} (grant no. PID2023-146453NB-100, PI: Beck).

The paper is based on spectroscopic observations made with the Southern
African Large Telescope (SALT). Polish participation in SALT is funded
by grant No. MEiN 2021/WK/01. This study uses observations collected at the European Organisation for Astronomical Research in the Southern Hemisphere under ESO programs 097.D-0338 and 099.D-0647. 

\section*{Data Availability}
Our photometric observations are available as supplementary material to this article. Additional photometric data used are accessible from the websites of the surveys. Optical spectra are available in the SALT archive, and infrared spectra are accessible in the ESO archive.



\bibliographystyle{mnras}
\bibliography{mnras_template} 



\appendix

\begin{figure}
\section{Additional photometry}
\includegraphics[width=\columnwidth]{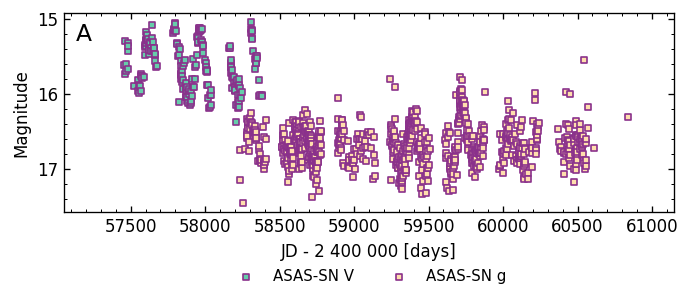}
\includegraphics[width=\columnwidth]{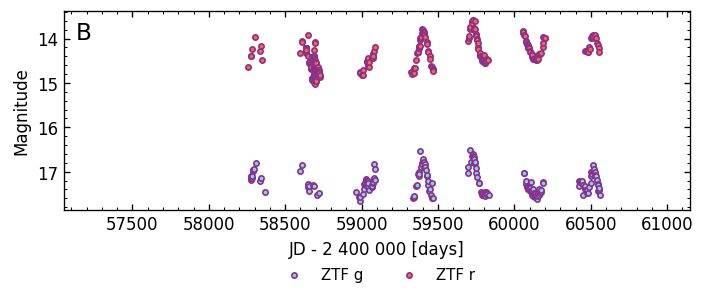}
\includegraphics[width=\columnwidth]{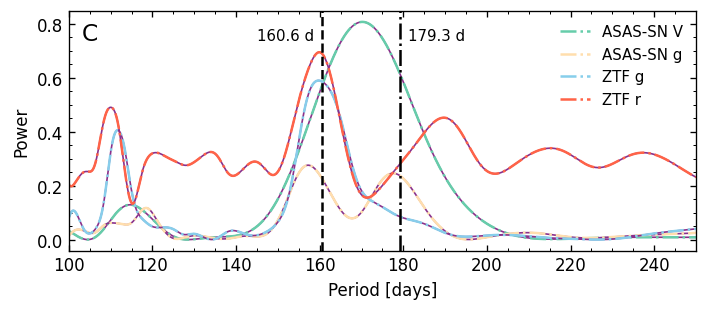}
    \caption{Light curves and periodograms of additional photometric data. \textbf{A:} Light curves from ASAS-SN survey in $V$ and $g$ filter. Light from both stars is blended in these datasets. \textbf{B:} ZTF $g$ and $r$ light curves of the blended sources. \textbf{C}: Lomb-Scargle periodograms of ASAS-SN and ZTF data, same as shown in Fig. \ref{fig:periodogram} for other datasets. Vertical lines indicate prominent periods identified from OGLE data of the individual stars. Additional peaks near 111 and 120 days are yearly aliases of the true periods.}
    \label{fig:phot_app}
\end{figure}

\newpage

\begin{table}
\section*{Supporting information}
\raggedright Supplementary data are available online.\\
	\centering
	\caption{Our photometry of Terz V 2513 in \textit{V} and \textit{I} filters. }
	\label{tab:our_photometry}
\begin{tabular}{rcccc}
\hline
JD         & \textit{V} {[}mag{]} & $\sigma_V$ {[}mag{]} & $I_c$ {[}mag{]} & $\sigma_{I_c}$ {[}mag{]} \\
\hline
2457200.33 & 16.15       & 0.25             & 12.04        & 0.03              \\
2457216.36 &             &                  & 12.10        & 0.03              \\
2457240.32 & 16.10       & 0.25             & 12.03        & 0.03              \\
2457255.30 & 16.22       & 0.25             & 12.11        & 0.03              \\
2457268.30 & 16.23       & 0.25             & 12.22        & 0.04              \\
2457426.60 & 16.01       & 0.25             & 12.03        & 0.03             \\
... & ...       & ...             & ...        & ...             \\
\hline
\end{tabular}
\end{table}


\bsp	
\label{lastpage}
\end{document}